\documentclass[reprint,amsmath,amssymb,aps,prb,superscriptaddress,longbibliography]{revtex4-1}
\usepackage{array}
\usepackage{bm}
\usepackage{color}
\usepackage{float}
\usepackage{graphicx}
\usepackage{hyperref}
\hypersetup{colorlinks=true,allcolors=blue}
\usepackage{natbib}
\usepackage{newtxtext}
\usepackage{newtxmath}
\usepackage[caption=false]{subfig}
\begin{document}

\title{Magnetic Field Induced Competing Phases in Spin-Orbital Entangled Kitaev Magnets}

\author{Li Ern Chern}
\affiliation{Department of Physics, University of Toronto, Toronto, Ontario M5S 1A7, Canada}

\author{Ryui Kaneko}
\affiliation{Institute for Solid State Physics, University of Tokyo, Kashiwa, Chiba 277-8581, Japan}

\author{Hyun-Yong Lee}
\affiliation{Institute for Solid State Physics, University of Tokyo, Kashiwa, Chiba 277-8581, Japan}

\author{Yong Baek Kim}
\affiliation{Department of Physics, University of Toronto, Toronto, Ontario M5S 1A7, Canada}
\affiliation{Perimeter Institute for Theoretical Physics, Waterloo, Ontario N2L 2Y5, Canada}
\affiliation{Canadian Institute for Advanced Research/Quantum Materials Program, Toronto, Ontario M5G 1Z8, Canada}
\affiliation{School of Physics, Korea Institute for Advanced Study, Seoul 130-722, Korea}
%\email[]{}
%\thanks{}


%\date{\today}

\begin{abstract}
There has been a great interest in magnetic field induced quantum spin liquids in Kitaev magnets after the discovery of neutron scattering continuum and half quantized thermal Hall conductivity in the material $\alpha$-RuCl$_3$. In this work, we provide a semiclassical analysis of the relevant theoretical models on large system sizes, and compare the results to previous studies on quantum models with small system sizes. We find a series of competing magnetic orders with fairly large unit cells at intermediate magnetic fields, which are most likely missed by previous approaches. We show that quantum fluctuations are typically strong in these large unit cell orders, while their magnetic excitations may resemble a scattering continuum and give rise to a large thermal Hall conductivity. Our work provides an important basis for a thorough investigation of emergent spin liquids and competing phases in Kitaev magnets.
\end{abstract}

\pacs{}
%\keywords{}

\maketitle


\noindent \textbf{\large Introduction} \\
Discovery of quantum spin liquids\cite{RevModPhys.89.025003,1903.08081} with emergent quasiparticles has been an important subject in modern condensed matter physics. This serves as an ultimate test of our understanding of highly quantum entangled phases in interacting electron systems. Recent research has invested tremendous effort on a number of materials with strong spin-orbit coupling\cite{annurev-conmatphys-020911-125138,annurev-conmatphys-031115-011319}, which leads to intriguing bond dependent exchange interactions between spin-orbital entangled pseudospin-$1/2$ moments. These studies are largely motivated by the exact solution of the Kitaev honeycomb model\cite{KITAEV20062}. The Kitaev interaction is naturally present in the systems with $4d/5d$ transition metal elements\cite{PhysRevLett.102.017205}, such as honeycomb/hyperhoneycomb iridates\cite{PhysRevLett.105.027204,Katukuri_2014,PhysRevLett.114.077202} and $\alpha$-RuCl$_3$\cite{PhysRevB.90.041112}. However, other exchange interactions are present too\cite{PhysRevLett.112.077204}, which often lead to magnetically ordered ground states instead of the desired quantum spin liquid\cite{PhysRevB.83.220403,PhysRevB.85.180403,PhysRevB.91.144420,PhysRevB.92.235119}. Hence much effort has been spent to suppress the magnetic orders and gain access to the possible spin liquid phases. \\

Over the past few years, great experimental progress has been achieved in $\alpha$-RuCl$_3$. At zero magnetic field, this material orders magnetically in the zigzag (ZZ) order\cite{PhysRevB.91.144420,PhysRevB.92.235119}. Upon the application of an external field, neutron scattering experiments\cite{PhysRevLett.120.077203,s41535-018-0079-2,1903.00056} find an intermediate window of fields before the system enters the polarized state, where sharp magnon modes are absent, but a scattering continuum appears instead. Under a $[111]$ field (perpendicular to the honeycomb plane), the measured thermal Hall conductivity above the ordering temperature $T_\mathrm{N} \approx 7 \, \mathrm{K}$ follows the predicted trend of itinerant Majorana fermions in the pure Kitaev model\cite{PhysRevLett.120.217205}. When the field is tilted away from the $[111]$ direction by $45$\textdegree\, and $60$\textdegree, half quantized thermal Hall conductivity is observed\cite{s41586-018-0274-0}. These observations raise the hope that the paramagnetic state in the intermediate field regime may be the sought-after chiral spin liquid with Majorana edge modes. \\

Theoretical models for $\alpha$-RuCl$_3$ include substantial Kitaev and symmetric anisotropic $\Gamma$ interactions, both strongly dependent on the bond directions, with additional small exchanges such as the neareast neighbor Heisenberg $J$, the third nearest-neighbor Heisenberg $J_3$, and the anisotropic $\Gamma'$\cite{PhysRevB.93.155143,PhysRevB.93.214431,PhysRevB.96.115103}, on the honeycomb lattice. Previous analyses are largely done on quantum models with small system sizes (typically a $24$-site cluster) via exact diagonalization (ED) \cite{s41535-018-0095-2,PhysRevB.99.064425,1901.09131,1901.09943,1904.01025} or in quasi-one dimensional limit via density matrix renormalization group (DMRG)\cite{PhysRevB.97.075126,1901.09131,1901.09943}, with varying degree of complexity. For example, a recent work\cite{1901.09943} on the $K \Gamma \Gamma'$ model in an external magnetic field suggests that it allows an intermediate spin liquid phase continuously connected to the pure Kitaev model between the low field ZZ order and high field polarized state. \\

\begin{figure}
\includegraphics[scale=0.3]{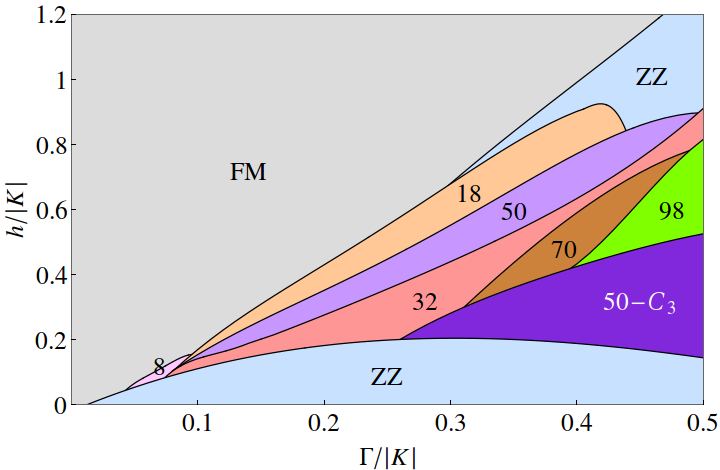}
\caption{\label{KGGpphase}Classical phase diagram of the $K \Gamma \Gamma'$ honeycomb model in a $[111]$ magnetic field $h$. Ferromagnetic Kitaev interaction $K=-1$ is assumed. $\Gamma'$ is fixed to be $-0.02$. Each of the integers indicates the total number of sublattices in a unit cell of the corresponding magnetic order.}
\end{figure}

In this article, we investigate the possible competing phases in the classical $K \Gamma \Gamma'$ model under a $[111]$ magnetic field for large system sizes. The purpose is to critically examine what kind of competing phases may be present and how these phases may be related to potential spin liquids in the quantum model. Rather surprisingly, we find a series of competing magnetic orders with large unit cells in the intermediate field regime. In particular, in the $K \Gamma \Gamma'$ model with small $\Gamma'$, the ground state in the zero field limit is the ZZ order, which is consistent with previous experiments and theoretical calculations. Upon increasing the field, the ZZ order is replaced by a series of magnetically ordered phases with $8$, $18$, $32$, $50$, $70$ and $98$-site unit cells before the system enters the polarized state (see Fig.~\ref{KGGpphase}). Hence the magnetic field reveals a series of competing orders, which form an intermediate region in the phase diagram. Most of these large unit cell orders had not been identified in previous works. \\

We compute the zero point quantum fluctuations for these magnetic orders and estimate the reduction of the size of the local moments. We find that quantum fluctuations are strong in the large unit cell orders so that the renormalized local moment is only about $50 \%$ of the full magnitude on average. The flat and dense spin wave spectra in the large unit cell orders, in particular the $70$ and $98$-site orders, essentially look like continua of spin excitations. Furthermore, we calculate the thermal Hall conductivity due to magnons in some of the large unit cell orders and find that it is as large as that observed experimentally at low temperatures. While strong quantum fluctuations are present and hence it is likely for the series of competing phases to turn into spin liquids in the quantum limit, it is also evident that previous theoretical studies on quantum models with small system sizes\cite{PhysRevB.97.075126,s41535-018-0095-2,PhysRevB.99.064425,1901.09131,1901.09943,1904.01025} cannot resolve many of these large unit cell orders. Therefore, in future analyses of such quantum models, it will be important to understand the role of quantum fluctuations in the large unit cell orders unveiled in the current work. Our findings demonstrate the possibility of novel and exotic ordering patterns in spin-orbital entangled Kitaev magnets, which provide an important basis for further investigations of the origin and the nature of quantum spin liquids that they may host. \\

\begin{figure}
\includegraphics[scale=0.3]{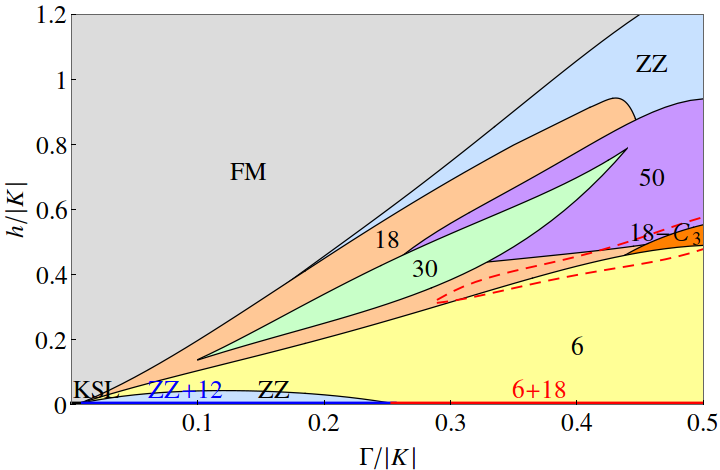}
\caption{\label{KGphase}Classical phase diagram of the $K \Gamma$ honeycomb model in a $[111]$ magnetic field $h$. Ferromagnetic Kitaev interaction $K=-1$ is assumed. Each of the integers indicates the total number of sublattices in a unit cell of the corresponding magnetic order, while $+$ indicates degeneracy. KSL denotes the extensively degenerate manifold of the Kitaev model, which only exists in the Kitaev limit $(\Gamma,h)=(0,0)$. The ground state in the parameter region enclosed by the red dashed line is likely an incommensurate order which exhibits domains of $18$ and $18$-$C_3$.}
\end{figure}

\noindent \textbf{\large Results} \\
\textbf{Model.} We investigate the nearest neighbor $K \Gamma \Gamma'$ model on the honeycomb lattice in a $[111]$ magnetic field $\mathbf{h}$,
\begin{equation} \label{KGGphamiltonian}
\begin{aligned}[b]
& H = \sum_{\lambda=x,y,z} \sum_{\langle ij \rangle \in \lambda} \left[ K S_i^\lambda S_j^\lambda + \Gamma ( S_i^\mu S_j^\nu + S_i^\nu S_j^\mu ) \right. \\
& + \left. \Gamma' ( S_i^\mu S_j^\lambda + S_i^\lambda S_j^\mu + S_i^\nu S_j^\lambda + S_i^\lambda S_j^\nu ) \right] - \mathbf{h} \cdot \sum_i \mathbf{S}_i ,
\end{aligned}
\end{equation}
where $K$ is the Kitaev interaction, $\Gamma$ and $\Gamma'$ are off diagonal spin exchanges, $(\lambda,\mu,\nu)$ is a cyclic permutation of $(x,y,z)$ and the field $\mathbf{h}=h (1,1,1) / \sqrt{3}$. We have also assumed an isotropic $g$ tensor. In \eqref{KGGphamiltonian}, $h$ actually carries a factor of $S$ but for notational simplicity we will just write $h$ in units of the Kitaev interaction, for instance $h=0.1 \lvert K \rvert$ instead of $h=0.1 \lvert K \rvert S$, in the rest of this article. \\

In the experimentally relevant parameter regime, $K<0$ and $\Gamma>0$ are large while $\Gamma'<0$ is small. In contrast to many of the previous studies\cite{PhysRevB.97.075126,s41535-018-0095-2,PhysRevB.99.064425, 1901.09943,1901.09131,1903.10026,1904.01025}, we investigate the classical limit of this model, that is, by treating the spins $\mathbf{S}_i=(S_i^x,S_i^y,S_i^z)$ in \eqref{KGGphamiltonian} as three dimensional vectors of fixed magnitude $\lvert \mathbf{S}_i \rvert = S$ for all $i$. We use simulated annealing to determine the ground state spin configuration of the system. Details of the simulated annealing calculation can be found in the Methods section. \\

\noindent \textbf{Phase diagrams.} We first consider the $K \Gamma$ model by setting $\Gamma'=0$ in \eqref{KGGphamiltonian}, with a ferromagnetic Kitaev interaction $K=-1$. We explore $\Gamma \in [0,0.5], h \in [0,1.2]$ and map out the phase diagram, as shown in Fig.~\ref{KGphase}. Apart from the extensively degenerate Kitaev limit $(\Gamma,h)=(0,0)$, we find that the vast majority of the parameter space favors particular magnetic orders. All these ordered phases, except the zigzag (ZZ) order and the ferromagnet (FM), are labeled by the number of sites contained in their respective magnetic unit cells. \\

In the zero field limit, the degeneracy of the Kitaev manifold is lifted as the ZZ order and a $12$-site order (a $6$-site order and an $18$-site order) are selected at small (intermediate) $\Gamma$. These two phases have exactly the same energy at $h=0$, but the ZZ order or the $6$-site order is prefered once $h \neq 0$. However, the 18-site order reemerges at higher fields and replaces the 6-site order as the ground state. Tracing back to the parameter region with small $\Gamma$ and $h$, we see that the ZZ, $6$-site, $12$-site and $18$-site orders are continuously connected to the Kitaev limit. The $6$-site order (the $18$-site order) was first reported in Ref.~\onlinecite{1807.06192} (Ref.~\onlinecite{PhysRevLett.117.277202}) and termed the X phase (the diluted star phase). At sufficiently large values of $\Gamma$ and $h$, even larger cluster ordering patterns like the $30$-site and $50$-site orders are stabilized. There is also an 18-site order with $C_3$ symmetry, which we label by $18$-$C_3$ to distinguish it from the previous $18$-site order as they are described by different arrangements of spins on the honeycomb lattice. \\

Next, we set $\Gamma'=-0.02$ and map out the phase diagram within the same ranges of $\Gamma$ and $h$, as shown in Fig.~\ref{KGGpphase}. The addition of such a small $\Gamma'$ term to the $K \Gamma$ model alters the phase diagram quite significantly. The degenerate manifold in the Kitaev limit and its neighborhood are replaced by the FM phase. The ZZ order is stabilized over a large portion of the parameter space at zero\cite{1408.4811} and low fields. Once again, we find at intermediate fields several large cluster ordering patterns, a 32-site order, a 70-site order, a 98-site order and a 50-site order with $C_3$ symmetry which we label by $50$-$C_3$. Finally, the strong $\Gamma$ high field regime of the phase diagram displays some similarities to the $\Gamma'=0$ case, where the same $50$-site, $18$-site and ZZ order are the lowest energy spin configurations, before the system becomes a FM. \\

\begin{figure}
\subfloat[]{\label{magnetization}
\includegraphics[scale=0.33]{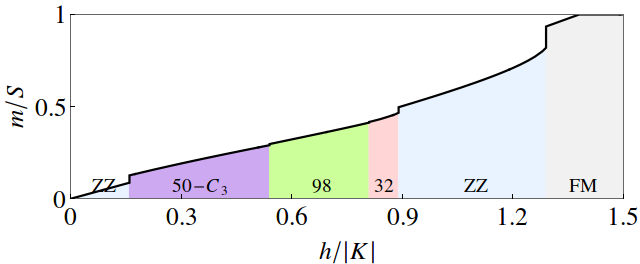}} \\
\subfloat[]{\label{reduction}
\includegraphics[scale=0.33]{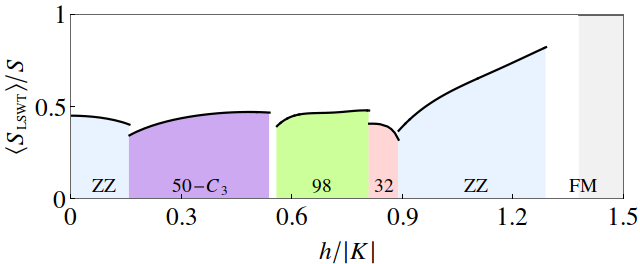}}
\caption{Magnetization and reduction of local moments. (a) The magnetization along the field direction, with the parametrization $(K,\Gamma,\Gamma')=(-1,0.5,-0.02)$ relevant to the material $\alpha$-RuCl$_3$. (b) The averaged renormalization of ordered moments when quantum fluctuations are taken into account via the linear spin wave theory with $S=1/2$, using the same parametrization as in (a), at zero temperature. Blank regions indicate that the spin wave Hamiltonian is not positive definite.}
\end{figure}

Details of the magnetic orders (the real space spin configurations and the static spin structure factors, etc.) that show up in the phase diagrams Figs.~\ref{KGphase} and \ref{KGGpphase}, from the four sublattice ZZ order to the 98-site order, can be found in the Supplementary Materials. We make some qualitative observations as follows. Firstly, stronger $\Gamma$ interaction stabilizes magnetic orders with larger unit cells. This is true for both zero and finite $\Gamma'$. We expect that ordering patterns with even larger unit cells than those mentioned above may appear if $\Gamma$ is further increased beyond $0.5$. Secondly, the large unit cell orders, like the $70$-site and $98$-site orders, are closely competing in the parameter region where they are stabilized. The difference in energy is typically $10^{-3}$ to $10^{-4}$ of the energies of these orders. Thirdly, the magnetic orders can be classified into two categories, one with an inversion symmetry and the other with a three fold rotational symmetry. The ZZ order and the magnetic orders labeled by numbers fall into the former, while the magnetic orders labeled by numbers appended with -$C_3$ fall into the latter. More details can be found in the Supplementary Materials. \\

\begin{figure}
\includegraphics[scale=0.3]{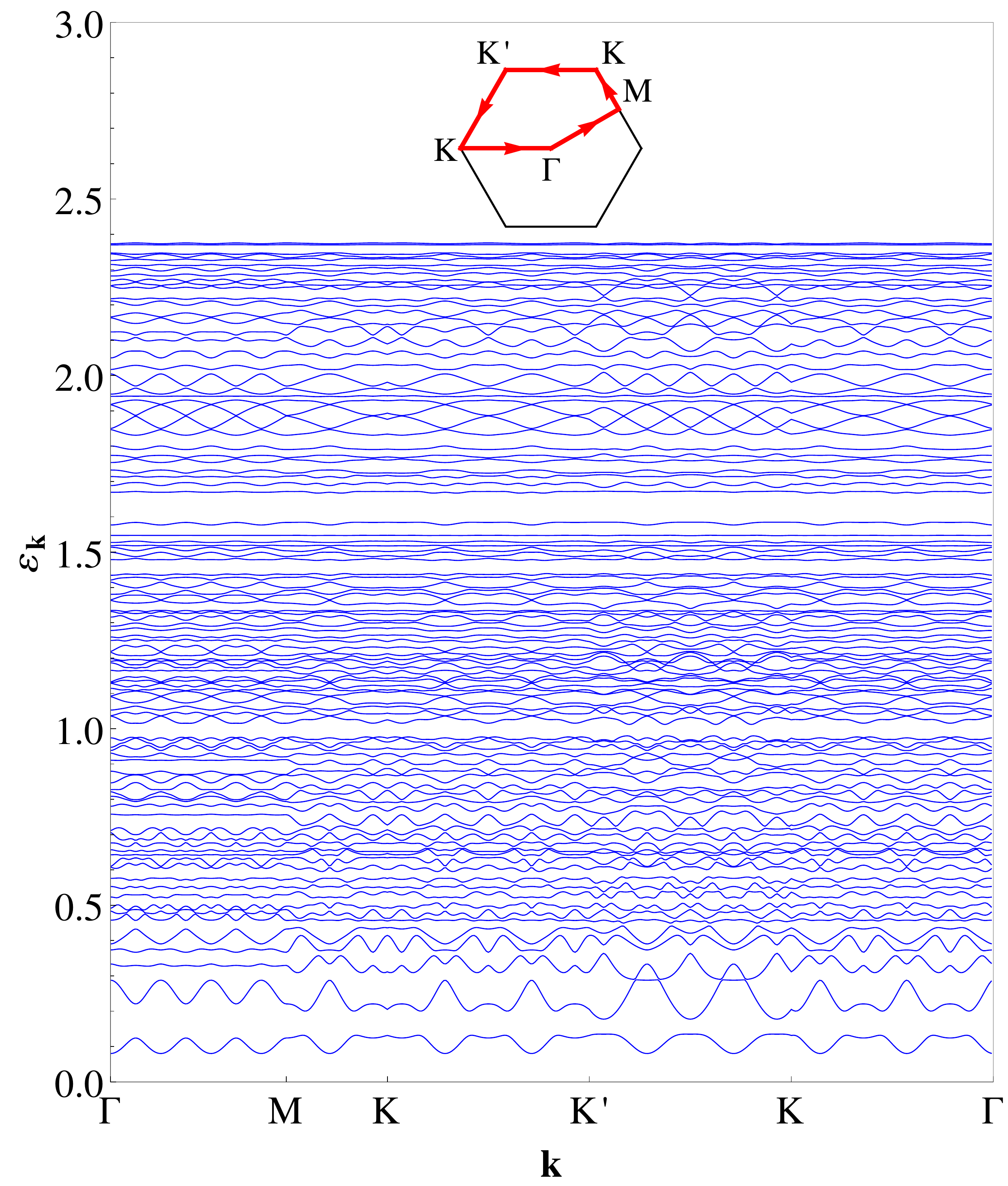}
\caption{\label{98spinwave}Spin wave dispersion of the $98$-site order. Plotted in units of $\lvert K \rvert S$, with the parametrization $(K,\Gamma,\Gamma')=(-1,0.5,-0.02)$ and at the field $h=0.6$. The inset shows the path travelled in the first Brillouin zone of the honeycomb lattice.}
\end{figure}

\begin{figure*}
\subfloat[]{\label{50C3hall}
\includegraphics[scale=0.3]{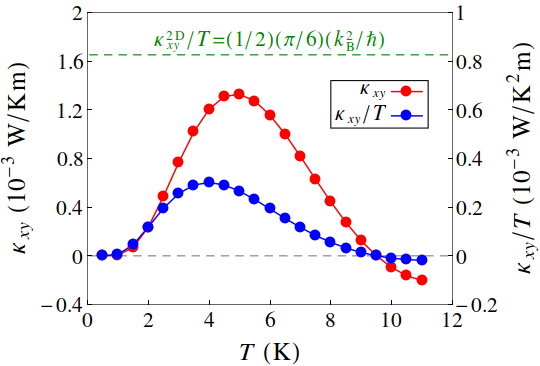}} \quad
\subfloat[]{\label{32hall}
\includegraphics[scale=0.3]{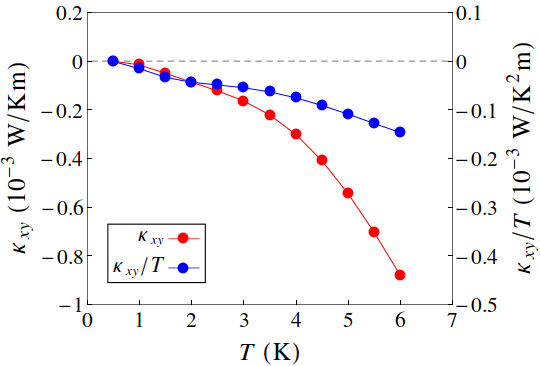}} \quad
\subfloat[]{\label{temperature}
\includegraphics[scale=0.3]{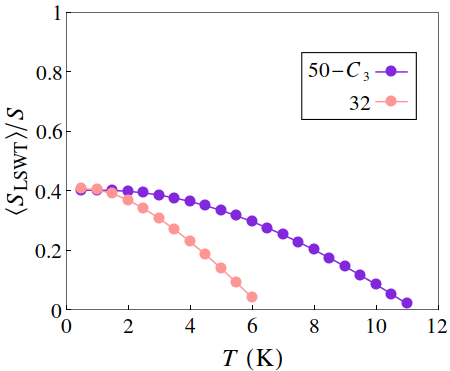}}
\caption{Thermal Hall conductivity due to magnons as a function of temperature. (a) The $50$-$C_3$ order at the field $h=0.23$ and (b) the $32$-site order at the field $h=0.82$, both with the parametrization $(K,\Gamma,\Gamma')=(-1,0.5,-0.02)$. We also indicate in (a) the half quantized thermal Hall conductivity $\kappa_{xy}^{2\mathrm{D}}=\kappa_{xy}d=(1/2)(\pi/6)(k_\mathrm{B}^2 / \hbar) T$, or $\kappa_{xy}/T \approx 0.826 \times 10^{-3} \, \mathrm{W}/\mathrm{K}^2 \mathrm{m}$ with the inter-plane distance $d=5.72$ \AA\cite{PhysRevLett.120.217205,s41586-018-0274-0}. (c) The averaged renormalization of ordered moments at finite temperatures calculated from the linear spin wave theory, for the two magnetic orders in (a) and (b). The Kitaev interaction is assumed to have a magnitude of $80 \, \mathrm{K}$.}
\end{figure*}

\noindent \textbf{Magnetization.} The proposed spin model for the material $\alpha$-RuCl$_3$ is parametrized by dominant $K$ and $\Gamma$ exchanges, with $K<0$ and $\Gamma \approx -K/2$, plus some small additional interactions like $\Gamma'$, $J$ and $J_3$, where $J$ ($J_3$) is the (third) nearest neighbor Heisenberg exchange\cite{PhysRevB.93.155143,PhysRevB.93.214431,PhysRevB.96.115103}. Therefore, in the phase diagram Fig.~\ref{KGGpphase} of the $K \Gamma \Gamma'$ model in a $[111]$ magnetic field, we take a cut along $\Gamma=0.5$ and plot the magnetization $m=\mathbf{S} \cdot \hat{\mathbf{h}}$ as a function of the field $h$, as shown in Fig.~\ref{magnetization}. The magnetization increases monotonically with the field, and jumps at the phase transitions. The discontinuities are not very obvious at the transitions between the large unit cell orders, but are significant when the system enters to (exit from) a large unit cell order from (to) ZZ, and from ZZ to FM. This suggests the difficulty of detecting phase transitions at intermediate fields by inspecting the magnetization, if they exist at all in the quantum model. \\

\noindent \textbf{Linear spin wave theory.} As a first approach to study the effect of quantum fluctuations on the classical orders, we apply the linear spin wave theory\cite{Jones_1987,PhysRevB.98.060412} to calculate the reduction of ordered moments in the zero temperature limit. For simplicity, we assume the same underlying magnetic orders, and do not consider how the classical phase diagram may be changed due to quantum correction to the energy because there are too many competing phases. Details of the linear spin wave calculation can be found in the Methods section. In Fig.~\ref{reduction}, we plot the average fraction of spins achieved in the linear spin wave theory with $S=1/2$ as a function of the field,
\begin{equation}
\frac{\langle S_\mathrm{LSWT} \rangle}{S} = 1-\frac{\sum_{i} \langle b_i^\dagger b_i \rangle / N_\mathrm{site}}{S} ,
\end{equation}
where $b_i^\dagger$ is the magnon creation operator at site $i$ and $N_\mathrm{site}$ is the total number of sites in the system. Blank regions indicate that the spin wave Hamiltonian is not positive definite at one or more momenta, i.e.~the lowest magnon band becomes gapless. At low and intermediate fields $h \lesssim 1$, the average reduction of ordered moments is about $50 \%$ of the full magnitude $S$, hinting at strong quantum fluctuations. At high fields $h \gtrsim 1$, $\langle S_\mathrm{LSWT} \rangle / S$ increase monotonically with $h$ in the ZZ phase, but the spin wave solution becomes unstable in the region $h \in [1.29,1.37]$, where the system is in the FM phase \textit{with the spins not completely aligning with the $[111]$ field} (see Fig.~\ref{magnetization}). Not only is this region likely to host quantum spin liquids, but it is also interesting from the classical aspect, which we will discuss in details later. Finally, for $h > 1.37$, the system enters the fully polarized state and $\langle S_\mathrm{LSWT} \rangle / S =1$ achieves saturation. \\

The spin wave dispersion of a (very) large unit cell order typically appears flat and dense. As an example, we show the spin wave dispersion of the $98$-site order along certain high symmetry directions in the first Brillouin zone of the honeycomb lattice in Fig.~\ref{98spinwave}. \\

\noindent \textbf{Thermal Hall conductivity.} We calculate the thermal Hall conductivity due to magnons\cite{PhysRevLett.104.066403,PhysRevB.89.054420,JPSJ.86.011010},
\begin{equation}\label{hallequation}
\kappa_{xy} = - \frac{k_\mathrm{B}^2 T}{\hbar V} \sum_\mathbf{k} \sum_{n=1}^\mathcal{N} \left \lbrace c_2 \left[ f_\mathrm{BE}(\varepsilon_{\mathbf{k},n}) \right] - \frac{\pi^2}{3} \right \rbrace \Omega_{\mathbf{k},n} .
\end{equation}
Details of the calculation can be found in the Methods section. Expressing the field in \eqref{KGGphamiltonian} as $\mathbf{h}=g \mu_\mathrm{B} \mu_0 \mathbf{H}$, assuming the $g$ factor $g=2.3$\cite{PhysRevB.98.060412,PhysRevB.98.094425} and the magnitude of the Kitaev interaction $\lvert K \rvert \approx 80 \, \mathrm{K}$\cite{PhysRevB.93.155143,PhysRevLett.120.217205}, $\mu_0 H=12 \, \mathrm{T}$ roughly corresponds to $h=0.23$. At this field and with the parametrization $(K,\Gamma,\Gamma')=(-1,0.5,-0.02)$, the system is in the $50$-$C_3$ order. We plot the thermal Hall conductivity $\kappa_{xy}$ as a function of temperature $T$, as shown in Fig.~\ref{50C3hall}. We show only data below $T_\mathrm{c} \approx 11 \, \mathrm{K}$, defined as the temperature at which $\langle S_\mathrm{LSWT} \rangle / S$ drops to zero, i.e.~the magnetic order is destroyed by thermal fluctuations (see Fig.~\ref{temperature}). We find that $\kappa_{xy}$ is close to zero but slightly negative at $10 \, \mathrm{K}$. It gradually develops a positive value as $T$ decreases, and peaks at $5 \, \mathrm{K}$ before diminishing again as $T \longrightarrow 0$. The maximum value of $\kappa_{xy}/T$ is about $0.3 \times 10^{-3} \, \mathrm{W} / \mathrm{K}^2 \mathrm{m}$, which is of the same order of magnitude as the half quantized value $0.826 \times 10^{-3} \, \mathrm{W}/\mathrm{K}^2 \mathrm{m}$ measured in Ref.~\onlinecite{s41586-018-0274-0}. \\

We also calculate the thermal Hall conductivity for another large unit cell order, the $32$-site order, at the field $h=0.82$ (which would roughly correspond to $\mu_0 H = 43 \, \mathrm{T}$) and with the same parametrization, as shown in Fig.~\ref{32hall}. This time $T_\mathrm{c} \approx 6 \, \mathrm{K}$ (see Fig.~\ref{temperature}) and $\kappa_{xy}$ is negative. Starting from zero temperature, $\kappa_{xy}$ grows in magnitude as $T$ increases, and reaches $- 0.9 \times 10^{-3} \, \mathrm{W} / \mathrm{Km}$ at $6 \, \mathrm{K}$. The trend and the magnitude of $\kappa_{xy}$ are similar to those reported in Ref.~\onlinecite{PhysRevLett.120.217205} at lower fields ($\mu_0 H=6$, $12$ and $15 \, \mathrm{T}$). Hence the opposite signs of $\kappa_{xy}$ may indicate the presence of different magnetic orders. \\

\noindent \textbf{Frustrated ferromagnet.} We notice that there is a window of $h$ where the system is a FM but not fully polarized, i.e.~the spins align uniformly but not in the direction of the $[111]$ field. Such a phase is also stabilized in the high field regime at other parametrizations $(K,\Gamma,\Gamma')$ including the $K \Gamma$ model, and the width of the window is usually larger for stronger $\Gamma$. In the following, we attempt to derive some analytical understanding of why this situation occurs. \\

\begin{figure}
\includegraphics[scale=0.3]{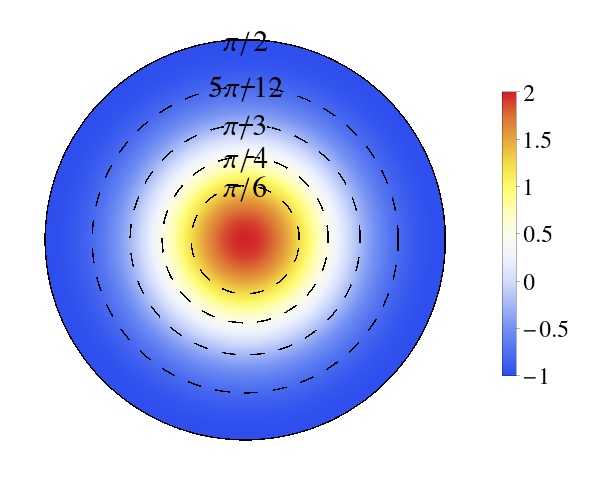}
\caption{\label{Genergy}Energy of the $\Gamma$ interaction in the FM phase. Plotted on the northern hemisphere, where the numbers on the circles indicate the zenith angle $\theta$ defined through $\sqrt{(S^a)^2+(S^b)^2}=\sin \theta, S^c=\cos \theta$. The north pole $\theta=0$ at the center corresponds to $S^c=1$, while the equator $\theta=\pi/2$ intersects the $ab$ plane. The energy possesses an azimuthal or $U(1)$ symmetry. The energy on the southern hemisphere is given by the same density plot.}
\end{figure}

We start from the $K \Gamma$ model with $K<0$, $\Gamma>0$, and $\Gamma'=0$ in \eqref{KGGphamiltonian}. Assuming a FM state, that is, $\mathbf{S}_i=\mathbf{S}$ for all sites $i$, the Hamiltonian reduces to
\begin{equation} \label{KGhamiltonianFM}
H = N \mathbf{S}^\mathrm{T} \left( H_K + H_\Gamma \right) \mathbf{S} - 2N \mathbf{h} \cdot \mathbf{S} ,
\end{equation}
with the matrices
\begin{equation} \label{KGmatrixFM}
H_K = \begin{pmatrix} K & 0 & 0 \\ 0 & K & 0 \\ 0 & 0 & K \end{pmatrix}, H_\Gamma = \begin{pmatrix} 0 & \Gamma & \Gamma \\ \Gamma & 0 & \Gamma \\ \Gamma & \Gamma & 0 \end{pmatrix} ,
\end{equation}
and $N$ being the total number of unit cells. The Kitaev interaction becomes ``isotropic'' in the FM state, behaving like the Heisenberg interaction. The $\Gamma$ interaction still appears quite anisotropic at this stage, but a change of basis will bring it to a simpler and more illuminating form. Switching from the cubic $xyz$ coordinates to the crystallographic $abc$ coordinates, where the $a$, $b$ and $c$ axes point in the directions $[11\bar{2}]$, $[\bar{1}10]$ and $[111]$ respectively\cite{PhysRevB.92.024413,PhysRevB.99.064425}, the spin is given by
\begin{equation}
\tilde{\mathbf{S}} = \begin{pmatrix} S^a \\ S^b \\ S^c \end{pmatrix} \\
= \begin{pmatrix}\frac{1}{\sqrt{6}} & \frac{1}{\sqrt{6}} & - \sqrt{\frac{2}{3}} \\ -\frac{1}{\sqrt{2}} & \frac{1}{\sqrt{2}} & 0 \\ \frac{1}{\sqrt{3}} & \frac{1}{\sqrt{3}} & \frac{1}{\sqrt{3}} \end{pmatrix} \begin{pmatrix} S^x \\ S^y \\ S^z \end{pmatrix} \\
= R \mathbf{S} .
\end{equation}
In the $abc$ basis, the Kitaev interaction $\tilde{H}_K = R H_K R^\mathrm{T} = H_K$ remains the same, while the $\Gamma$ interaction assumes the form of an $\mathrm{XXZ}$ model
\begin{equation}
\tilde{H}_\Gamma = R H_\Gamma R^\mathrm{T} = \begin{pmatrix} -\Gamma & 0 & 0 \\ 0 & -\Gamma & 0 \\ 0 & 0 & 2 \Gamma \end{pmatrix} .
\end{equation}
We can then analyze \eqref{KGhamiltonianFM} in the $abc$ basis term by term. It can be shown analytically that the energy of the classical Kitaev model is $ K \lvert \mathbf{S} \rvert^2 $ per unit cell\cite{PhysRevB.78.115116}. Thus any FM phase will minimize the energy of the $K$ term. On the other hand, the $\Gamma$ term attains maximum (minimum) when the spin points along (lies on) the $c$-axis ($ab$-plane). The energy profile of the $\Gamma$ term is shown in Fig.~\ref{Genergy}. \\

Suppose that the field $\tilde{\mathbf{h}}=(0,0,h)$ is along the $c$-axis or the $[111]$ direction. The $h$ term wants to align the spin with the $c$-axis, but this will be costly in energy for the $\Gamma$ term. The competition between $\Gamma$ and $h$ tilts the spin away from the $c$-axis. Therefore, such a FM phase can be stabilized between the fully polarized state and some other orders, e.g.~ZZ and $18$, in the high field regime. In contrast, if the field is along any of the in-plane directions, then all the $K$, $\Gamma$ and $h$ terms in \eqref{KGhamiltonianFM} can be minimized simultaneously. \\

Now let us consider the $K \Gamma \Gamma'$ model. A finite $\Gamma'$ term acts similarly as $\Gamma$. One can easily show that, assuming a FM state, $H_{\Gamma'}$ has the same structure as $H_\Gamma$ in \eqref{KGmatrixFM}. Thus a small $\Gamma'<0$ ($\Gamma'>0$) weakens (enhances) the effect of $\Gamma$. A similar FM but not fully polarized phase due to the presence of a large $\Gamma$ interation in the $J K \Gamma J_3$ model under a $[001]$ field was also found and discussed in Ref.~\onlinecite{PhysRevB.96.064430}. \\

\noindent \textbf{\large Discussion} \\
Classcially, the pure Kitaev model is extremely sensitive to an external magnetic field. It is polarized whenever the field $h \neq 0$. From the result of our simulated annealing calculation, a finite $\Gamma$ interaction on top of $K$ gives rise to a multitude of ordered phases, many of them possess fairly large magnetic unit cells, at finite fields. As $\Gamma$ increases, the window of these nontrivial magnetic orders becomes wider, and the system becomes polarized at greater value of $h$. Thus the combination of $\Gamma$ and $h$ effects like a prism that produce a rich and colorful phase diagram. Adding a small $\Gamma'$ term stabilizes even larger cluster magnetic orders at intermediate fields. We successfully demonstrate that the $K \Gamma \Gamma'$ honeycomb model is a playground for many exotic field induced magnetic orders, not simply the zigzag (ZZ) order and the polarized state as largely perceived in the past. \\

We discuss the implications of these large unit cell orders. First of all, the size of the system has to be sufficiently large to host them. If the system is smaller than or incommensurate with the magnetic order, the ground state spin configuration may appear like a disordered state. This calls for a serious reconsideration of the results from quantum calculations on small systems where finite size effect can be important, such as ED on the $24$-site cluster\cite{1901.09943} and iDMRG on the cylinder geometry\cite{PhysRevB.97.075126}, which report a quantum spin liquid ground state. The large unit cell magnetic orders found in this work cannot be captured by these and similar computations\cite{s41535-018-0095-2,PhysRevB.99.064425,1901.09131,1904.01025} on quantum models with small system sizes. Nevertheless, the possibility of a quantum spin liquid still exists, especially in the vicinity of the large unit cell orders where quantum fluctuations are strong. The large unit cell orders are very close in energy in the parameter region where they are stabilized. In addition, the average spin wave correction to the ordered moments in the large unit cell orders for a representative parametrization of $\alpha$-RuCl$_3$ is found to be more than $50 \%$. One can imagine that quantum fluctuations may melt these competing orders and promote a spin liquid state, but we will not know whether this is true until the magnetic orders are explicitly taken into account in the quantum model. If the large unit cell orders (partially) survive under quantum fluctuations, the magnon bands are typically flat and very close to each other such that they appear like the excitation continuum seen in inelastic neutron scattering experiments, which is often interpreted as fractionalized excitations in a quantum spin liquid\cite{nmat4604,Banerjee1055,PhysRevLett.120.077203,s41535-018-0079-2,1903.00056}. The resulting two magnon excitations will also form a very broad continuum at low energies. Moreover, we calculate the magnon thermal Hall conductivity for two of the large unit cell orders and show that it resembles the trend and/or the magnitude as that measured in experiments\cite{PhysRevLett.120.217205,s41586-018-0274-0} below the ordering temperature. In contrast, as computed in Ref.~\onlinecite{PhysRevB.98.060412}, the magnon thermal Hall conductivity in the ZZ order is in general quite small in magnitude. \\

We also discover the existence of a ferromagnetic (FM) but not fully polarized state at high fields in the $K \Gamma \Gamma'$ model with zero or small $\Gamma'$, which can be understood through the competition between the $\Gamma$ and $h$ terms. Here $K<0$ and $\Gamma>0$ are assumed. While the field always wants to orient the spins in its direction, the $\Gamma$ interaction is only minimized (maximized) when the spins are all lying on the $ab$-plane (pointing along the $c$-axis). This may explain why the system is more prone to polarization when the tilting angle of the field from the $[111]$ direction is larger. This also suggests that frustration is stronger (weaker) when the field is along (in) the $c$-axis ($ab$-plane). For instance, the simulated annealing calculation on the classical $J K \Gamma J_3$ honeycomb model in an in-plane field\cite{1807.06192} with the parametrization $(J,K,\Gamma,J_3)=(-0.035,-1,0.5,0.035)$ only yield the $6$-site order (termed the X phase) at intermediate fields, between the ZZ order at low fields and the polarized state at high fields, thus leading to a relatively simple phase diagram. \\

\noindent \textbf{\large Methods} \\
\textbf{Simulated annealing.} The simulated annealing calculation is performed on a honeycomb lattice with $L \times L$ unit cells (or $L \times L \times 2$ sites) with periodic boundary conditions. Most of the computations are done with $L=12,15,20$, but sometimes other $L$ is used when the ground state spin configuration is not obvious. The procedure of simulated annealing is outlined as follows. In the beginning, we generate a totally random spin configuration on the honeycomb lattice and define a ``temperature'' parameter $T$. We randomly select a site on the honeycomb lattice and propose a random orientation for the spin on that site. Next, we calculate the difference in energy and accept the change with the probability $\min \lbrace 1, \exp(-\Delta E/T) \rbrace$. This step is repeated for $\sim 10^7$ times at a fixed $T$, which is then decreased gradually. Once $T<T_\mathrm{c}$ for some critical temperature $\sim 10^{-8} \lvert K \rvert$, we update the spin at site $i$ deterministically by aligning it in the direction of the local field\cite{PhysRevLett.117.277202} defined as
\begin{equation} \label{localfield}
\mathbf{B}_i = -\sum_j H_{ij} \mathbf{S}_j + \mathbf{h} ,
\end{equation}
where $H_{ij}$ is the three dimensional matrix that encodes the interaction between the spins at $i$ and $j$. \\

If the sublattice structure of a magnetic order is known, we can carry out the above procedure for a small number of spins and calculate the energy to very high precision. This allows us to better determine the phase boundary between competing orders. \\

\noindent \textbf{Linear spin wave theory.} The content in this section is mainly derived from Ref.~\onlinecite{Jones_1987}. For each sublattice $i$ in the magnetic unit cell, we first choose a local coordinates system in which the spin $\mathbf{S}_i$ aligns in the $z$-direction. The amounts to a change of basis characterized by the rotation matrix
\begin{equation}
R_i = \begin{pmatrix} \cos \theta_i \cos \phi_i & -\sin \phi_i & \sin \theta_i\cos \phi_i \\ \cos \theta_i \sin \phi_i & \cos \phi_i & \sin \theta_i \sin \phi_i \\ - \sin \theta_i & 0 & \cos \theta_i \end{pmatrix} ,
\end{equation}
where $\theta_i$ and $\phi_i$ are the two angles parametrizing the orientation of $\mathbf{S}_i$ in the cubic $xyz$ coordinates, $(S_i^x,S_i^y,S_i^z) = S (\sin \theta_i \cos \phi_i, \sin \theta_i \sin \phi_i, \cos \theta_i)$. The third column of $R_i$ is precisely $\mathbf{S}_i$ up to the factor $S$, while the first and second columns are chosen such that the three columns are mutually orthonormal and satisfy the right hand rule. We define $\mathbf{S}_i=R_i \tilde{\mathbf{S}}_i$. Classically, we have $\tilde{\mathbf{S}}_i = (0,0,S)$. Quantum effects on the ordered moments are introduced through spin wave excitations (magnons),
\begin{subequations}
\begin{align}
& \tilde{S}_i^z = S-b_i^\dagger b_i = S-n_i, \label{SzSW} \\
& \tilde{S}_i^x = \frac{\sqrt{2S-n_i} b_i + b_i^\dagger \sqrt{2S-n_i}}{2} \approx \sqrt{\frac{S}{2}} (b_i + b_i^\dagger), \label{SxSW} \\
& \tilde{S}_i^y = \frac{\sqrt{2S-n_i} b_i - b_i^\dagger \sqrt{2S-n_i}}{2i} \approx -i \sqrt{\frac{S}{2}} (b_i - b_i^\dagger), \label{SySW}
\end{align}
\end{subequations}
where we have used the linear spin wave approximation that neglects the third and higher order terms in $b_i$ in the series expansion of \eqref{SxSW} and \eqref{SySW}. Next, we rewrite the spin Hamiltonian as
\begin{equation} \label{hamiltonian}
H = \sum_{ij} \mathbf{S}_i^\mathrm{T} H_{ij} \mathbf{S}_j - \mathbf{h}^\mathrm{T} \sum_i \mathbf{S}_i = \sum_{ij} \tilde{\mathbf{S}}_i^\mathrm{T} \tilde{H}_{ij} \tilde{\mathbf{S}}_j - \sum_i \tilde{\mathbf{h}}_i^\mathrm{T} \tilde{\mathbf{S}}_i,
\end{equation}
where $\tilde{H}_{ij} = R_i^\mathrm{T} H_{ij} R_j$ and $\tilde{\mathbf{h}}_i = R_i \mathbf{h}$. Representing $\tilde{\mathbf{S}}_i$ using \eqref{SzSW}-\eqref{SySW}, keeping only terms quadratic in $b_i$, and performing a Fourier transform
\begin{equation} \label{fouriertransform}
b_{\mathbf{k},s} = \frac{1}{\sqrt{N}} \sum_i b_{i,s} e^{\mathbf{k} \cdot \mathbf{R}_i} ,
\end{equation}
where, from now on, $i$ denotes the position in the Bravais lattice defined by the translational symmetries of the magnetic order, $s$ denotes the sublattice in the magnetic unit cell, and $N$ is the total number of magnetic unit cells. We then obtain the spin wave Hamiltonian in momentum space
\begin{equation}
H = \sum_\mathbf{k} \Psi_\mathbf{k}^\dagger \mathrm{D}_\mathbf{k} \Psi_\mathbf{k} ,
\end{equation}
where $\Psi_\mathbf{k} = (b_{\mathbf{k},1}, \ldots, b_{\mathbf{k},\mathcal{N}}, b_{-\mathbf{k},1}^\dagger, \ldots, b_{-\mathbf{k},\mathcal{N}}^\dagger)$ and $\mathcal{N}$ is the total number of sublattices in the magnetic unit cell. $\mathrm{D}_\mathbf{k}$ is a $2 \mathcal{N}$ dimensional matrix of the form
\begin{equation}
\mathrm{D}_\mathbf{k} = \begin{pmatrix} \mathrm{A}_\mathbf{k} & \mathrm{B}_\mathbf{k} \\ \mathrm{B}_{-\mathbf{k}}^* & \mathrm{A}_{-\mathbf{k}}^\mathrm{T} \end{pmatrix} ,
\end{equation}
where $\mathrm{A}_\mathbf{k}$ and $\mathrm{B}_\mathbf{k}$ are $\mathcal{N}$ dimensional matrices. To obtain the spin wave dispersion, we diagonalize $\mathrm{D}_\mathbf{k}$ by a Bogoliubov transformation in order to to preserve the canonical commutation relation of the bosons,
\begin{equation}\label{bogoliubov}
\mathrm{T}_\mathbf{k}^\dagger \mathrm{D}_\mathbf{k} \mathrm{T}_\mathbf{k} = \mathcal{E}_\mathbf{k}, \mathrm{T}_\mathbf{k} \sigma^3 \mathrm{T}_\mathbf{k}^\dagger = \sigma^3,
\end{equation}
where $\mathcal{E}_\mathbf{k}=\mathrm{diag}(\varepsilon_{\mathbf{k},1},\ldots,\varepsilon_{\mathbf{k},\mathcal{N}},\varepsilon_{-\mathbf{k},1},\ldots,\varepsilon_{-\mathbf{k},\mathcal{N}})$ and $\sigma^3$ is a diagonal matrix with the first $\mathcal{N}$ entries equal to $1$ and the last $\mathcal{N}$ entries equal to $-1$. The average reduction of ordered moments \eqref{SzSW} at temperature $T$ can be calculated from the matrix elements of the Bogoliubov transformation,
\begin{equation}
\begin{aligned}[b]
& \frac{1}{N_\mathrm{site}} \sum_{is} \langle b_{is}^\dagger b_{is} \rangle = \frac{1}{N \mathcal{N}} \sum_{\mathbf{k}} \sum_{m,n=1}^\mathcal{N} \left \lbrace \mathrm{T}_\mathbf{k}^*(m,n) \mathrm{T}_\mathbf{k} (m,n) f_\mathrm{BE} (\varepsilon_{\mathbf{k},n}) \right. \\
& \left. + \mathrm{T}_\mathbf{k}^*(m,n+\mathcal{N}) \mathrm{T}_\mathbf{k} (m,n+\mathcal{N}) \left[ 1+f_\mathrm{BE} (\varepsilon_{-\mathbf{k},n}) \right] \right \rbrace ,
\end{aligned}
\end{equation}
where $f_\mathrm{BE}$ is the Bose-Einstein distribution,
\begin{equation}\label{boseeinstein}
f_\mathrm{BE} (\varepsilon) = \frac{1}{e^{\varepsilon/T}-1} .
\end{equation}

\noindent \textbf{Thermal Hall conductivity.} We explain the various symbols that appear in the formula \eqref{hallequation} for the calculation of the thermal Hall conductivity\cite{PhysRevB.89.054420,JPSJ.86.011010}. $n$ is the magnon band index that runs from $1$ to $\mathcal{N}$. The function $c_2$ is given by
\begin{equation}
\begin{aligned}[b]
c_2(x) &= \int_0^x \mathrm{d}t \left( \ln \frac{1+t}{t} \right)^2 \\
&= (1+x) \left( \ln \frac{1+x}{x} \right)^2 - (\ln x)^2-2\mathrm{Li}_2(-x) ,
\end{aligned}
\end{equation}
where $\mathrm{Li}_2$ is the dilogarithm. $f_\mathrm{BE}$ is the Bose-Einstein distribution as defined in \eqref{boseeinstein}. $\Omega_{\mathbf{k},n}$ is the Berry curvature defined as
\begin{equation}
\Omega_{\mathbf{k},n} = i \epsilon_{\mu \nu} \left[ \sigma^3 \frac{\partial \mathrm{T}_\mathbf{k}^\dagger}{\partial k_\mu} \sigma^3 \frac{\partial \mathrm{T}_\mathbf{k}}{\partial k_\nu} \right]_{nn} ,
\end{equation}
where $\sigma^3$ and $\mathrm{T}_\mathbf{k}$ are defined as in \eqref{bogoliubov}. For the calculation of the total volume $V$ of the system, we use the inter-plane distance $5.72$ \AA \, between the honeycomb layers in $\alpha$-RuCl$_3$\cite{PhysRevLett.120.217205,s41586-018-0274-0}. The exact value of the in-plane lattice constant does not enter the calculation explicitly because, while $1/V$ contribute two inverse factors of it, $\Omega_{\mathbf{k},n}$ contribute two factors, so they cancel out. When performing the summation over momenta in \eqref{hallequation}, we partition the first Brillouin zone (of the magnetic order) evenly such that it contains a total number of $L \times L$ $\mathbf{k}$ points. We check the convergence of $\kappa_{xy}$ with increasing $L$ up to $L=800$. We also ensure that the Chern number of each magnon band,
\begin{equation}
C_n = \frac{1}{2 \pi} \sum_\mathbf{k} \frac{(2 \pi)^2}{A} \Omega_{\mathbf{k},n} ,
\end{equation}
where $A$ is the total area of the system, converges to an integer with increasing $L$. \\

\noindent \textbf{Data availability.} The data that support the findings of this study are available from the corresponding author upon reasonable request. \\

\bibliography{reference190520}

\vspace{12 pt}
\noindent \textbf{\large Acknowledgement} \\
We thank Hae-Young Kee, Jacob Gordon, Jonathan Cookmeyer and Kyusung Hwang for useful discussions. We also thank Lukas Janssen for bringing our attention to Ref.~\onlinecite{PhysRevB.96.064430}. L.E.C. is grateful to Andrei Catuneanu, Panagiotis Peter Stavropoulos and Sopheak Sorn for teaching him the particulars of the simulated annealing calculation. R.K. was supported by MEXT as ``Priority Issue on Post-K computer'' (Creation of New Functional Devices and High-Performance Materials to Support Next-Generation Industries). H.-Y.L. was supported by MEXT as ``Exploratory Challenge on Post-K computer'' (Frontiers of Basic Science: Challenging the Limits). Y.B.K. was supported by the Killam Research Fellowship from the Killam Foundation, the NSERC of Canada and the Center for Quantum Materials at the University of Toronto. Most of the computations were performed on the Niagara supercomputer at the SciNet HPC Consortium\cite{1742-6596-256-1-012026}. SciNet is funded by: the Canada Foundation for Innovation; the Government of Ontario; Ontario Research Fund - Research Excellence; and the University of Toronto. Part of the computations were performed at the Supercomputer Center, ISSP, University of Tokyo. \\

\noindent \textbf{\large Author contributions} \\
L.E.C. did the main calculation. R.K. and H.-Y.L. crosschecked the result. Y.B.K. conceived and supervised the study. All authors contributed to the writing of the manuscript. \\

\noindent \textbf{\large Additional information} \\
\textbf{Supplementary Materials} accompany this paper at \textit{website}. \\

\noindent \textbf{Competing interests.} The authors declare no competing interests.

\clearpage

\onecolumngrid

\begin{center}
\textbf{\large Supplementary Materials: \\ Magnetic Field Induced Competing Phases in Spin-Orbital Entangled Kitaev Magnets}
\end{center}
\begin{center}
Li Ern Chern$^1$, Ryui Kaneko$^2$, Hyun-Yong Lee$^2$, Yong Baek Kim$^{1,3,4,5}$
\end{center}
\begin{center}
{\small
\textit{$^1$Department of Physics, University of Toronto, Toronto, Ontario M5S 1A7, Canada} \\
\textit{$^2$Institute for Solid State Physics, University of Tokyo, Kashiwa, Chiba 277-8581, Japan} \\
\textit{$^3$Perimeter Institute for Theoretical Physics, Waterloo, Ontario N2L 2Y5, Canada} \\
\textit{$^4$Canadian Institute for Advanced Research/Quantum Materials Program, Toronto, Ontario M5G 1Z8, Canada} \\
\textit{$^5$School of Physics, Korea Institute for Advanced Study, Seoul 130-722, Korea}
}
\end{center}

\setcounter{equation}{0}
\setcounter{figure}{0}
\setcounter{table}{0}
\setcounter{page}{1}

\renewcommand{\thesection}{S\arabic{section}}
\renewcommand{\theequation}{S\arabic{equation}}
\renewcommand{\thefigure}{S\arabic{figure}}
\renewcommand{\thetable}{S\arabic{table}}

\section*{\label{MOdetail}Details of the Magnetic Orders}
\begin{figure}[h]
\subfloat[]{\label{ZZansatz} \raisebox{24 pt}{\includegraphics[scale=0.32]{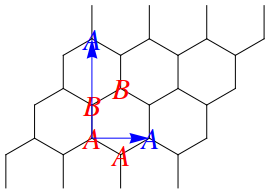}}}
\subfloat[]{\label{ZZconfiguration} \includegraphics[scale=0.28]{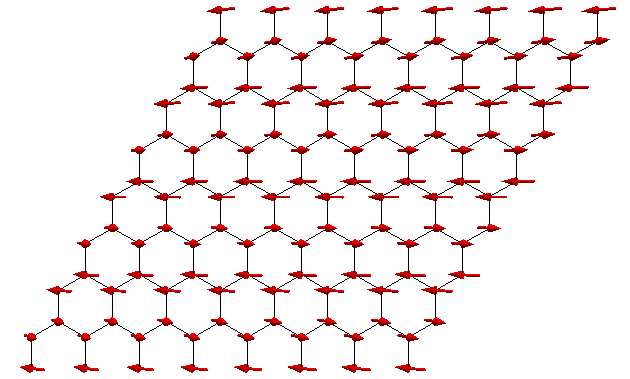}} \quad
\subfloat[]{\label{ZZstructurefactor} \includegraphics[scale=0.3]{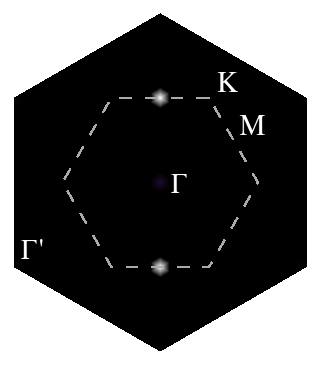}} \qquad
\subfloat[]{\label{structurefactorlegend} \raisebox{5 pt}{\includegraphics[scale=0.3]{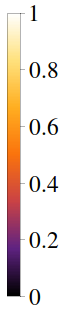}}}
\caption{\label{MOstart}(a) The sublattice structure in the magnetic unit cell of the ZZ order and the primitive vectors of the Bravais lattice it defines. (b) The spin configuration of the ZZ order shown on a finite segment of the honeycomb lattice, with the parameters $K=-1$, $\Gamma=0.2$, $\Gamma'=-0.02$ and $h=0.1$. The $[111]$ direction ($c$-axis) is perpendicular to the honeycomb plane. (c) The static spin structure factor of the ZZ order, which peaks at $\mathbf{k}=\mathrm{M}$, with the same parameters in (b). The intensity is normalized such that the maximum is $1$. (d) The color legend indicating the intensity of the structure factor. This is same for all the subsequent plots, so it will not be shown again.}
\end{figure}

\begin{figure}[h]
\subfloat[]{\label{6ansatz} \raisebox{36 pt}{\includegraphics[scale=0.32]{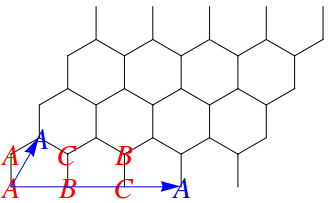}}}
\subfloat[]{\label{6configuration} \includegraphics[scale=0.28]{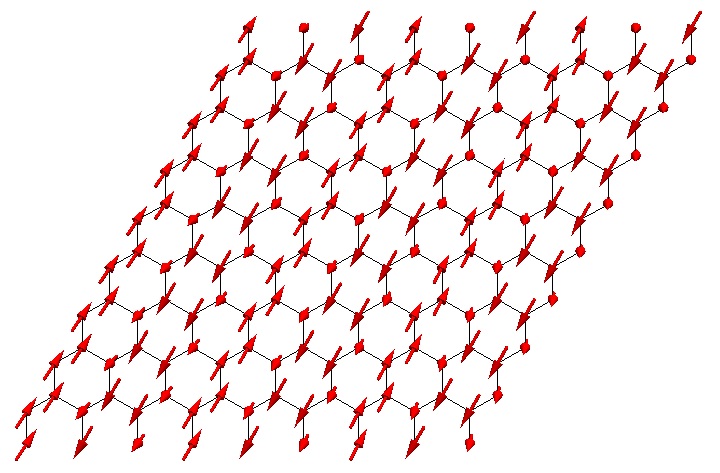}} \quad
\subfloat[]{\label{6structurefactor} \raisebox{8 pt}{\includegraphics[scale=0.3]{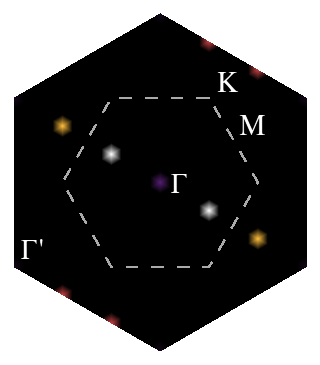}}}
\caption{(a) The sublattice structure in the magnetic unit cell of the $6$-site order and the primitive vectors of the Bravais lattice it defines. (b) The spin configuration of the $6$-site order shown on a finite segment of the honeycomb lattice, with the parameters $K=-1$, $\Gamma=0.3$, $\Gamma'=0$ and $h=0.12$. (c) The static spin structure factor of the $6$-site order, which peaks at $\mathbf{k}=\frac{2}{3} \mathrm{M}$, with the same parameters in (b).}
\end{figure}

\begin{figure}[h]
\subfloat[]{\label{8ansatz} \raisebox{24 pt}{\includegraphics[scale=0.32]{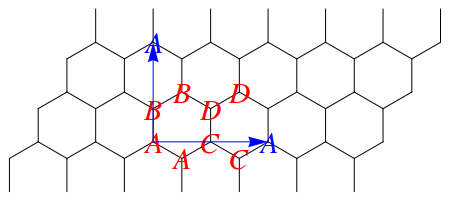}}}
\subfloat[]{\label{8configuration} \includegraphics[scale=0.28]{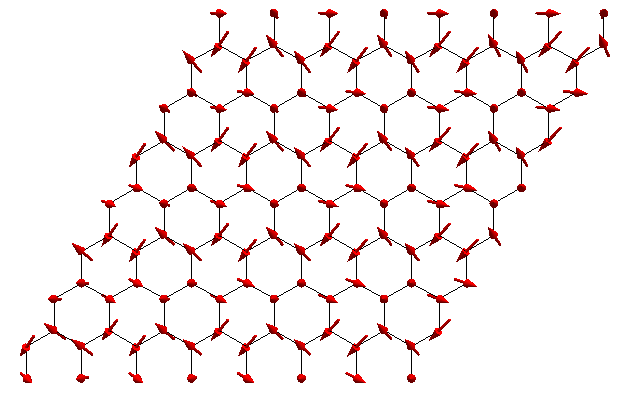}} \quad
\subfloat[]{\label{8structurefactor} \includegraphics[scale=0.3]{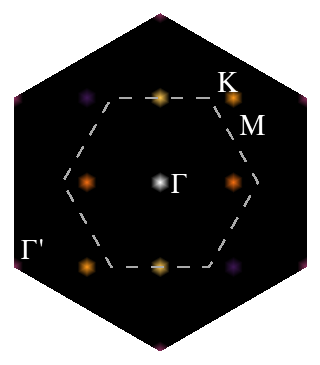}}
\caption{(a) The sublattice structure in the magnetic unit cell of the $8$-site order and the primitive vectors of the Bravais lattice it defines. (b) The spin configuration of the $8$-site order shown on a finite segment of the honeycomb lattice, with the parameters $K=-1$, $\Gamma=0.06$, $\Gamma'=-0.02$ and $h=0.08$. (c) The static spin structure factor of the $8$-site order, which peaks at $\mathbf{k}=\Gamma, \mathrm{M}, \frac{3}{4} \mathrm{K}, \frac{3}{4} \mathrm{K}'$ (the latter two only along one of the three fold directions), with the same parameters in (b).}
\end{figure}

\begin{figure}[h]
\subfloat[]{\label{12ansatz} \raisebox{24 pt}{\includegraphics[scale=0.32]{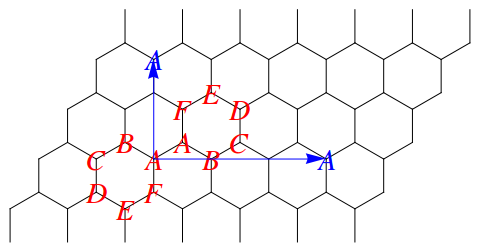}}}
\subfloat[]{\label{12configuration} \includegraphics[scale=0.28]{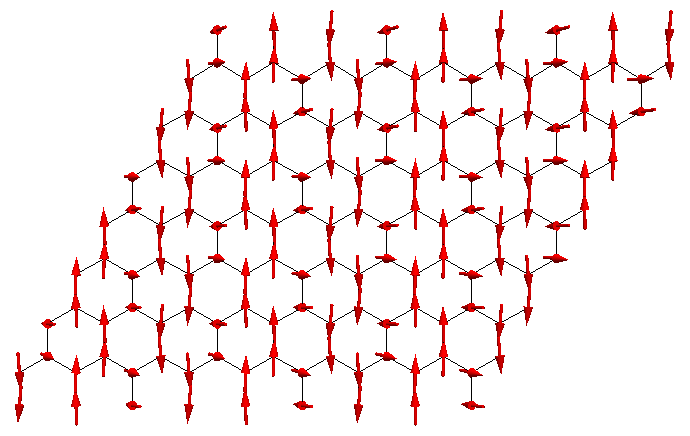}} \quad
\subfloat[]{\label{12structurefactor} \raisebox{3 pt}{\includegraphics[scale=0.3]{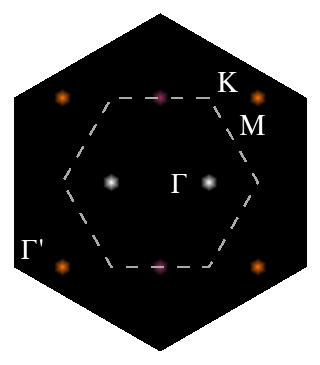}}}
\caption{(a) The sublattice structure in the magnetic unit cell of the $12$-site order and the primitive vectors of the Bravais lattice it defines. (b) The spin configuration of the $12$-site order shown on a finite segment of the honeycomb lattice, with the parameters $K=-1$, $\Gamma=0.1$, $\Gamma'=0$ and $h=0$. (c) The static spin structure factor of the $12$-site order, which peaks at $\mathbf{k}=\frac{1}{2} \mathrm{K}, \frac{1}{2} \mathrm{K}'$ (only along one of the three fold directions), with the same parameters in (b).}
\end{figure}

\begin{figure}[h]
\subfloat[]{\label{18lowansatz} \raisebox{24 pt}{\includegraphics[scale=0.32]{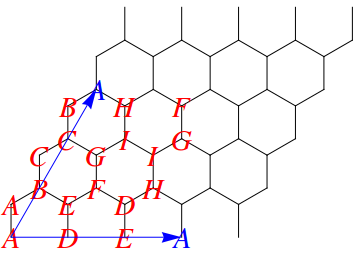}}}
\subfloat[]{\label{18lowconfiguration} \includegraphics[scale=0.28]{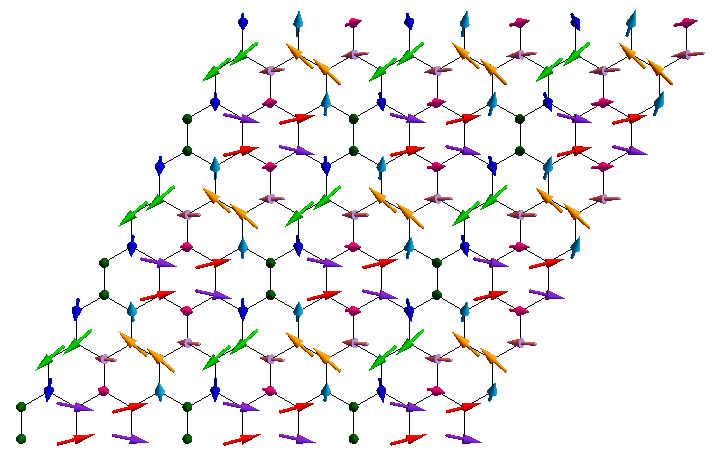}} \quad
\subfloat[]{\label{18lowstructurefactor} \raisebox{5 pt}{\includegraphics[scale=0.3]{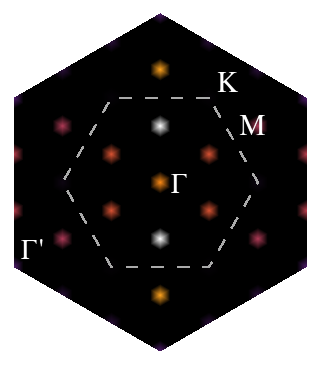}}}
\caption{(a) The sublattice structure in the magnetic unit cell of the $18$-site order and the primitive vectors of the Bravais lattice it defines. (b) The spin configuration of the $18$-site order at low fields shown on a finite segment of the honeycomb lattice, with the parameters $K=-1$, $\Gamma=0.3$, $\Gamma'=0$ and $h=0.36$. The spins with the same color are related by the inversion $\mathcal{I}$ in real space, and they point along the same direction in spin space. The centers of inversion are located in the middle of the bonds shared by pairs of identical spins. (c) The static spin structure factor of the $18$-site order at low fields, which peaks at $\mathbf{k}=\Gamma, \frac{2}{3} \mathrm{M}_i$, with the same parameters in (b). The intensities along the three fold directions are different, one direction is higher, while others are lower.}
\end{figure}

\begin{figure}[h]
\subfloat[]{\label{18highansatz} \raisebox{24 pt}{\includegraphics[scale=0.32]{18ansatz.png}}}
\subfloat[]{\label{18highconfiguration} \includegraphics[scale=0.28]{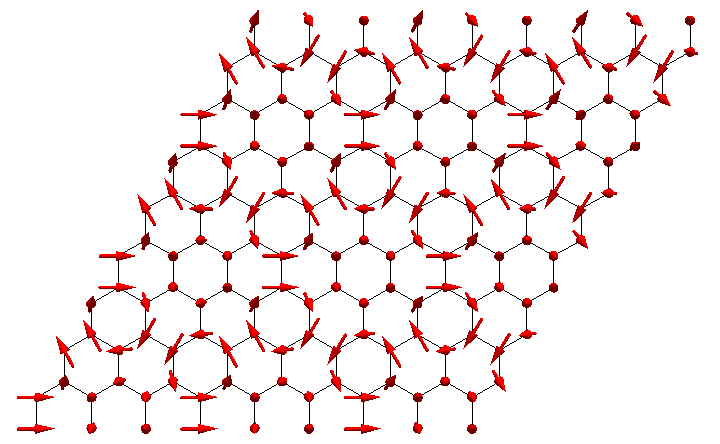}} \quad
\subfloat[]{\label{18highstructurefactor} \includegraphics[scale=0.3]{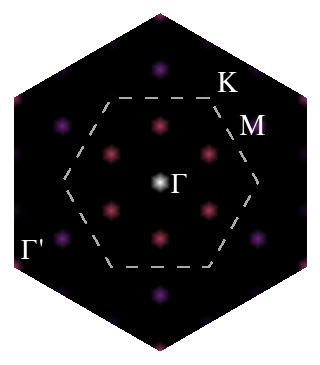}}
\caption{(a) The sublattice structure in the magnetic unit cell of the $18$-site order and the primitive vectors of the Bravais lattice it defines, same as Fig.~\ref{18lowansatz}. (b) The spin configuration of the $18$-site order at high fields shown on a finite segment of the honeycomb lattice, with the parameters $K=-1$, $\Gamma=0.3$, $\Gamma'=0$ and $h=0.6$. On top of the inversion $\mathcal{I}$ symmetry, it is $C_3$ symmetric about the center of a honeycomb plaquette, unlike the magnetic orders in Class $C_3$, which is $C_3$ symmetric about a site. (c) The static spin structure factor of the $18$-site order at high fields, which peaks at $\mathbf{k}=\Gamma, \frac{2}{3} \mathrm{M}_i$, with the same parameters in (b). The intensities along the three fold directions are same. Comparing to Fig.~\ref{18lowstructurefactor}, the intensities at $\frac{2}{3} \mathrm{M}_i$ become lower, while the intensity at $\Gamma$ becomes higher, as more spins are aligning towards the field direction.}
\end{figure}

\begin{figure}[h]
\subfloat[]{\label{18C3ansatz} \raisebox{24 pt}{\includegraphics[scale=0.32]{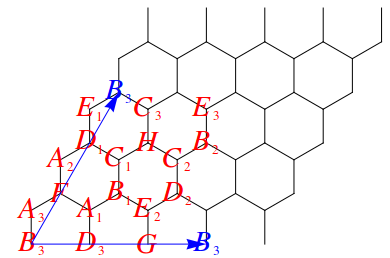}}}
\subfloat[]{\label{18C3configuration} \includegraphics[scale=0.28]{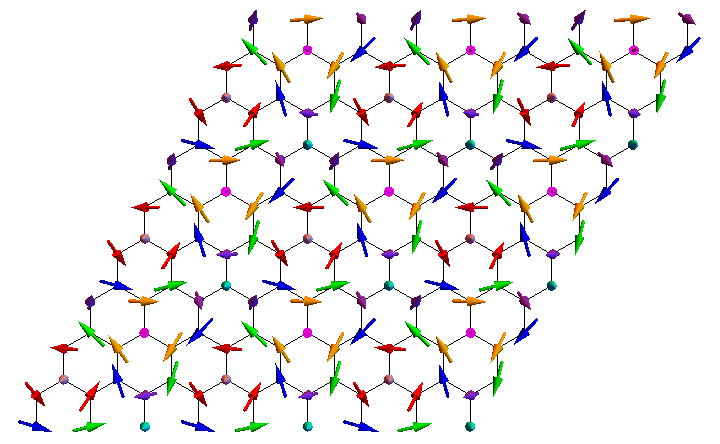}} \quad
\subfloat[]{\label{18C3structurefactor} \raisebox{5 pt}{\includegraphics[scale=0.3]{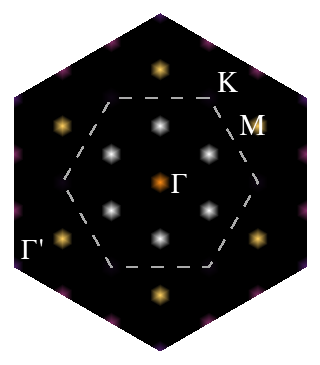}}}
\caption{(a) The sublattice structure in the magnetic unit cell of the $18$-$C_3$ order and the primitive vectors of the Bravais lattice it defines. The subscripts on the alphabets indicate the degree of permutation of the $x$, $y$ and $z$ spin components. The alphabets without a subscript indicate the centers of rotation. Spins \textit{F} and \textit{G} point in the $[111]$ direction, while spin \textit{H} points in the $[\bar{1}\bar{1}\bar{1}]$ direction. (b) The spin configuration of the $18$-$C_3$ order shown on a finite segment of the honeycomb lattice, with the parameters $K=-1$, $\Gamma=0.5$, $\Gamma'=0$ and $h=0.52$. The spins with the same color are related by the three fold rotation $C_3$ in both the real space and the spin space. The spin configuration is invariant under $C_3$. (c) The static spin structure factor of the $18$-$C_3$ order, which peaks at $\mathbf{k}=\Gamma$, $\frac{2}{3} \mathrm{M}_i$, with the same parameters in (b). The intensities along the three fold directions are same.}
\end{figure}

\begin{figure}[h]
\subfloat[]{\label{30ansatz} \raisebox{16 pt}{\includegraphics[scale=0.32]{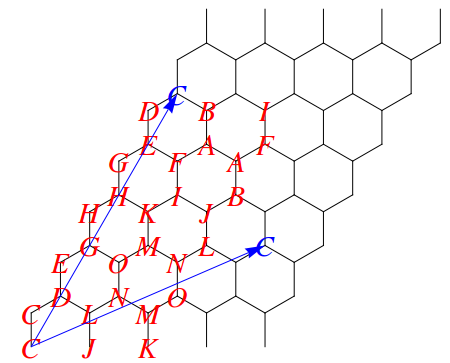}}}
\subfloat[]{\label{30configuration} \includegraphics[scale=0.28]{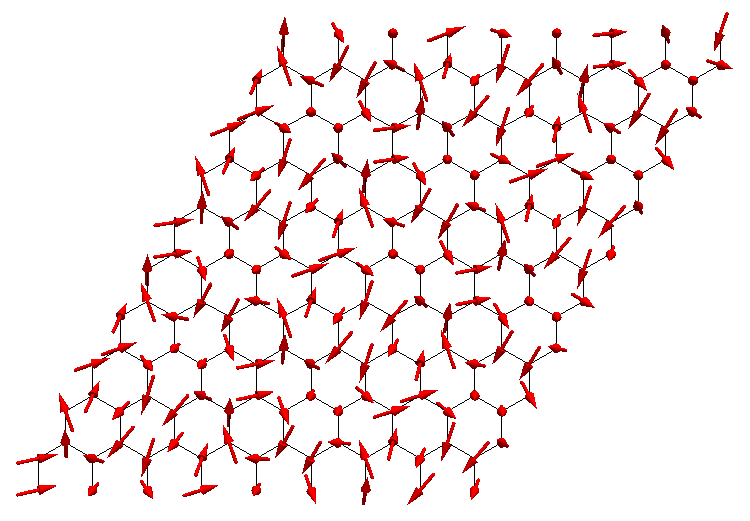}} \quad
\subfloat[]{\label{30structurefactor} \raisebox{14 pt}{ \includegraphics[scale=0.3]{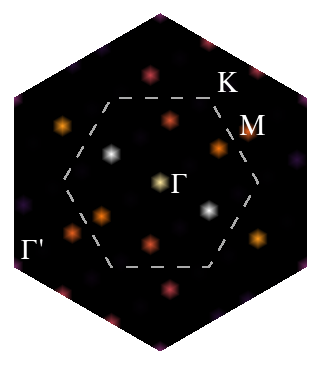}}}
\caption{(a) The sublattice structure in the magnetic unit cell of the $30$-site order and the primitive vectors of the Bravais lattice it defines. (b) The spin configuration of the $30$-site order shown on a finite segment of the honeycomb lattice, with the parameters $K=-1$, $\Gamma=0.3$, $\Gamma'=0$ and $h=0.42$. (c) The static spin structure factor of the $30$-site order, which peaks at $\mathbf{k}=\Gamma, \frac{2}{3} \mathrm{M}_1, \frac{4}{5} \mathrm{M}_2, \frac{2}{3} \mathrm{M}_1 + \frac{4}{5} \mathrm{M}_2$, with the same parameters in (b).}
\end{figure}

\begin{figure}[h]
\subfloat[]{\label{32ansatz} \raisebox{24 pt}{\includegraphics[scale=0.32]{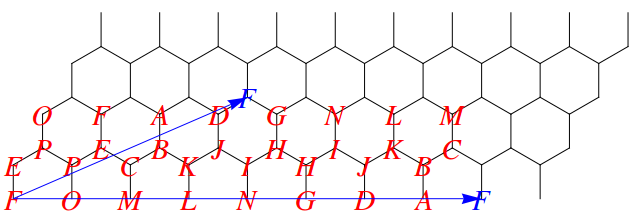}}}
\subfloat[]{\label{32configuration} \includegraphics[scale=0.28]{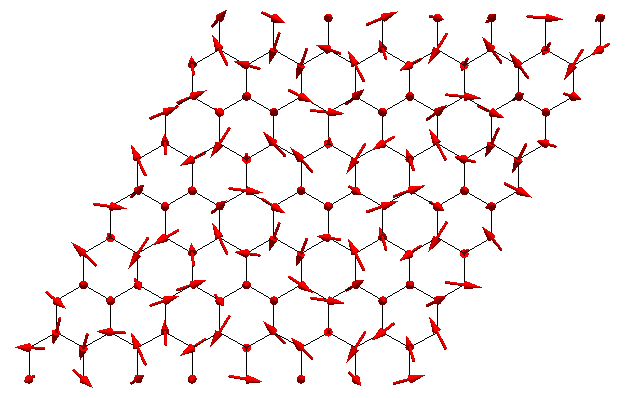}} \quad
\subfloat[]{\label{32structurefactor} \includegraphics[scale=0.3]{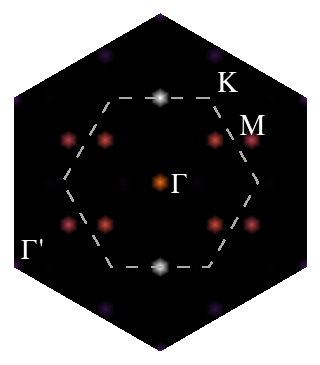}}
\caption{(a) The sublattice structure in the magnetic unit cell of the $32$-site order and the primitive vectors of the Bravais lattice it defines. (b) The spin configuration of the $32$-site order shown on a finite segment of the honeycomb lattice, with the parameters $K=-1$, $\Gamma=0.3$, $\Gamma'=-0.02$ and $h=0.36$. (c) The static spin structure factor of the $32$-site order, which peaks at $\mathbf{k}=\Gamma, \mathrm{M}, \pm (\frac{3}{8} \mathbf{b}_1 + \frac{7}{16} \mathbf{b}_2), \pm (\frac{3}{8} \mathbf{b}_1 - \frac{1}{16} \mathbf{b}_2)$, with the same parameters in (b).}
\end{figure}

\begin{figure}[h]
\subfloat[]{\label{50lowansatz} \raisebox{16 pt}{\includegraphics[scale=0.32]{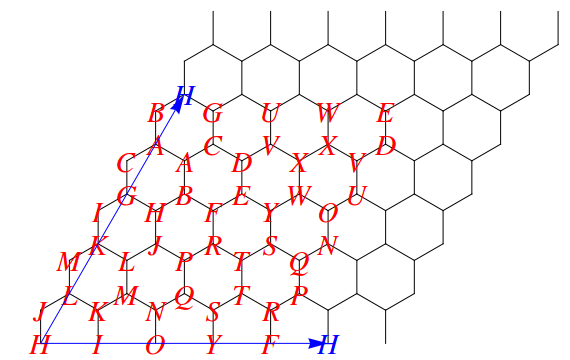}}}
\subfloat[]{\label{50lowconfiguration} \includegraphics[scale=0.28]{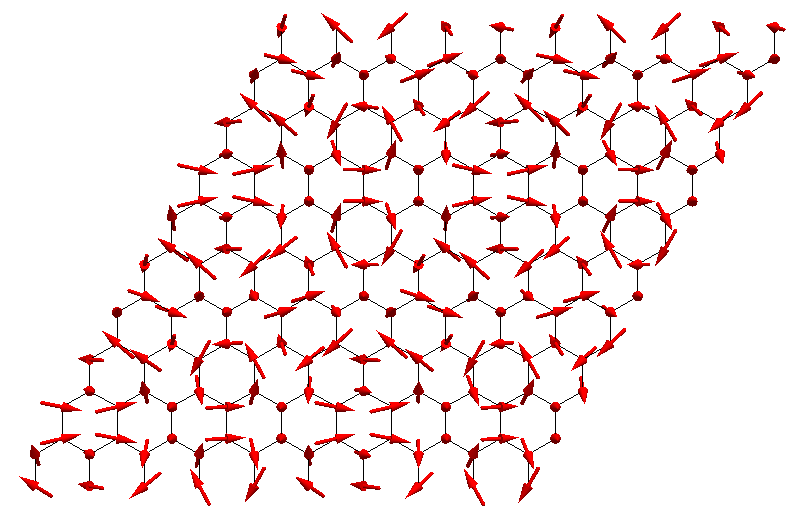}}
\subfloat[]{\label{50lowstructurefactor} \raisebox{14 pt}{\includegraphics[scale=0.3]{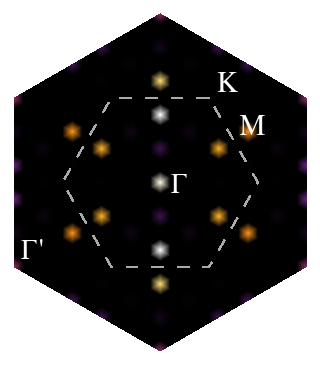}}}
\caption{(a) The sublattice structure in the magnetic unit cell of the $50$-site order and the primitive vectors of the Bravais lattice it defines. (b) The spin configuration of the $50$-site order at low fields shown on a finite segment of the honeycomb lattice, with the parameters $K=-1$, $\Gamma=0.4$, $\Gamma'=0$ and $h=0.54$. Comparing to Fig.~\ref{18lowconfiguration}, it looks like an augmented version of the $18$-site order at low fields. (c) The static spin structure factor of the $50$-site order at low fields, which peaks at $\mathbf{k}=\Gamma, \frac{4}{5} \mathrm{M}_i$, with the same parameters in (b). The intensities along the three fold directions are different, one direction is higher, while others are lower.}
\end{figure}

\begin{figure}[h]
\subfloat[]{\label{50highansatz} \raisebox{16 pt}{\includegraphics[scale=0.32]{50ansatz.png}}}
\subfloat[]{\label{50highconfiguration} \includegraphics[scale=0.28]{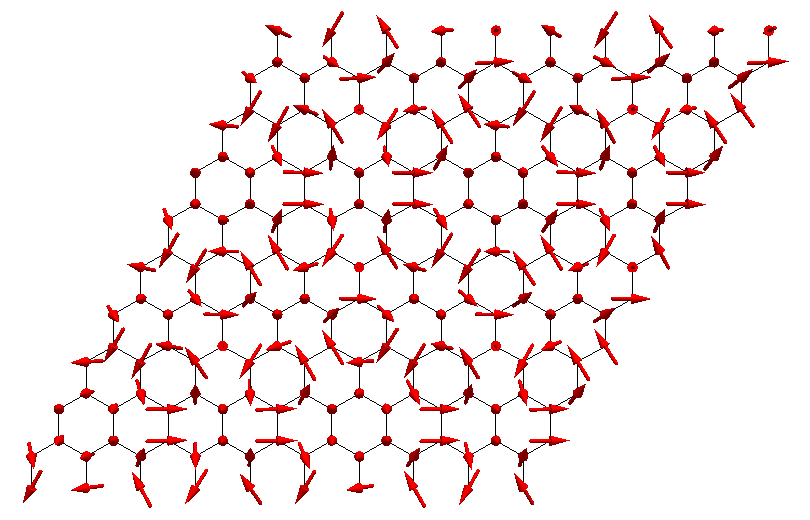}}
\subfloat[]{\label{50highstructurefactor} \raisebox{14 pt}{\includegraphics[scale=0.3]{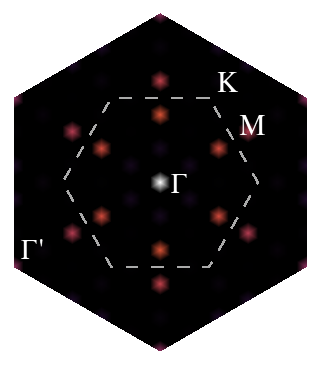}}}
\caption{(a) The sublattice structure in the magnetic unit cell of the $50$-site order and the primitive vectors of the Bravais lattice it defines, same as Fig.~\ref{50lowansatz}. (b) The spin configuration of the $50$-site order at high fields shown on a finite segment of the honeycomb lattice, with the parameters $K=-1$, $\Gamma=0.4$, $\Gamma'=0$ and $h=0.74$. On top of the inversion $\mathcal{I}$ symmetry, it is $C_3$ symmetric about the center of a honeycomb plaquette, unlike the magnetic orders in Class $C_3$, which is $C_3$ symmetric about a site. Comparing to Fig.~\ref{18highconfiguration}, it looks like an augmented version of the $18$-site order at high fields. (c) The static spin structure factor of the $50$-site order at high fields, which peaks at $\mathbf{k}=\Gamma, \frac{4}{5} \mathrm{M}_i$, with the same parameters in (b). The intensities along the three fold directions are same. Comparing to Fig.~\ref{50lowstructurefactor}, the intensities at $\frac{4}{5} \mathrm{M}_i$ become lower, while the intensity at $\Gamma$ becomes higher, as more spins are aligning towards the field direction.}
\end{figure}

\begin{figure}[h]
\subfloat[]{\label{50C3ansatz} \raisebox{16 pt}{\includegraphics[scale=0.32]{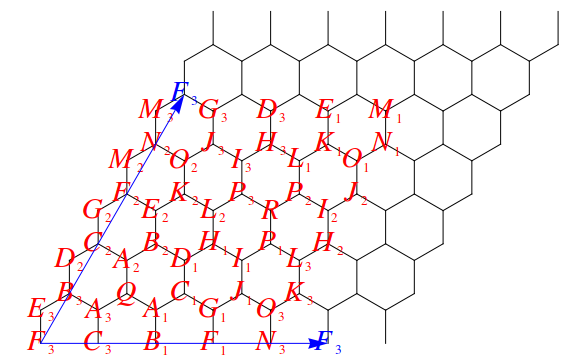}}}
\subfloat[]{\label{50C3configuration} \includegraphics[scale=0.28]{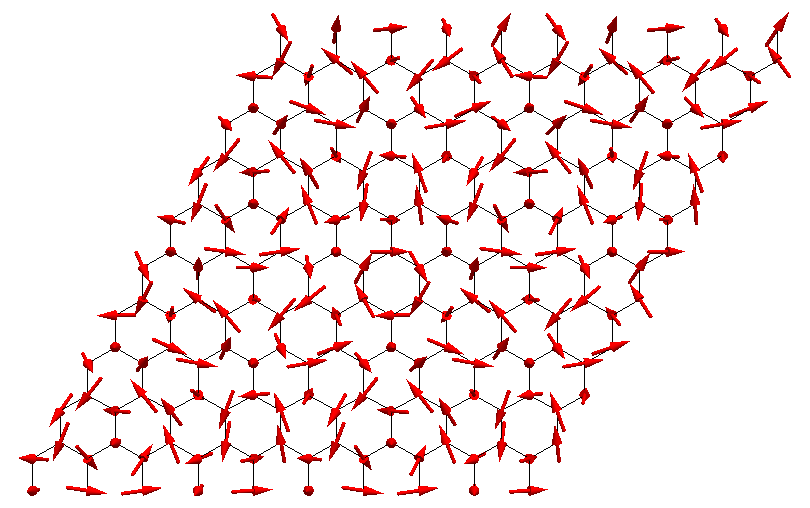}}
\subfloat[]{\label{50C3structurefactor} \raisebox{14 pt}{\includegraphics[scale=0.3]{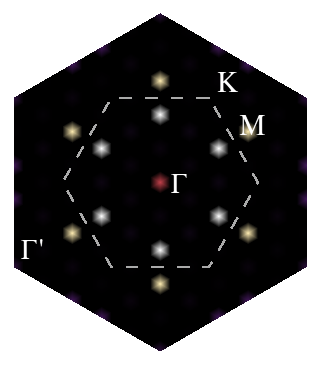}}}
\caption{(a) The sublattice structure in the magnetic unit cell of the $50$-$C_3$ order and the primitive vectors of the Bravais lattice it defines. The subscripts on the alphabets indicate the degree of permutation of the $x$, $y$ and $z$ spin components. The alphabets without a subscript indicate the centers of rotation. Spins \textit{Q} and \textit{R} point in the $[111]$ direction. (b) The spin configuration of the $50$-$C_3$ order shown on a finite segment of the honeycomb lattice, with the parameters $K=-1$, $\Gamma=0.5$, $\Gamma'=-0.02$ and $h=0.3$. Comparing to Fig.~\ref{18C3configuration}, it looks like an augmented version of the $18$-$C_3$ order. (c) The static spin structure factor of the $50$-$C_3$ order at high fields, which peaks at $\mathbf{k}=\Gamma, \frac{4}{5} \mathrm{M}_i$, with the same parameters in (b). The intensities along the three fold directions are same.}
\end{figure}

\begin{figure}[h]
\subfloat[]{\label{70ansatz} \includegraphics[scale=0.32]{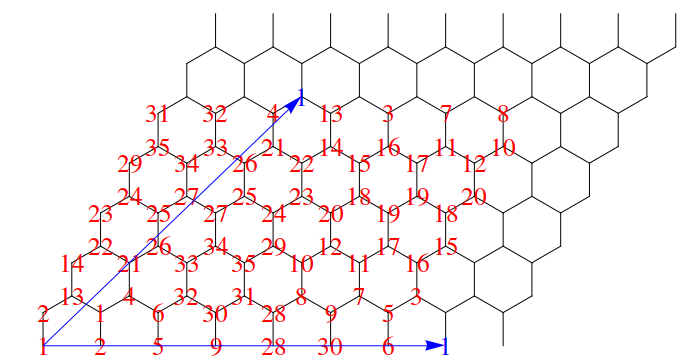}} \quad
\subfloat[]{\label{70structurefactor} \includegraphics[scale=0.3]{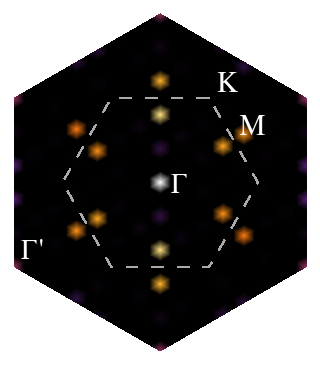}}
\caption{(a) The sublattice structure in the magnetic unit cell of the $70$-site order and the primitive vectors of the Bravais lattice it defines. For brievity, we do not show the real space spin configuration. (b) The static spin structure factor of the $70$-site order, which peaks at $\mathbf{k}=\Gamma, \frac{4}{5} \mathrm{M}_1, \frac{6}{7} \mathrm{M}_2, \frac{4}{5} \mathrm{M}_1+ \frac{6}{7} \mathrm{M}_2$, with the parameters $K=-1$, $\Gamma=0.4$, $\Gamma'=-0.02$ and $h=0.57$.}
\end{figure}

\begin{figure}[h]
\subfloat[]{\label{98ansatz} \includegraphics[scale=0.32]{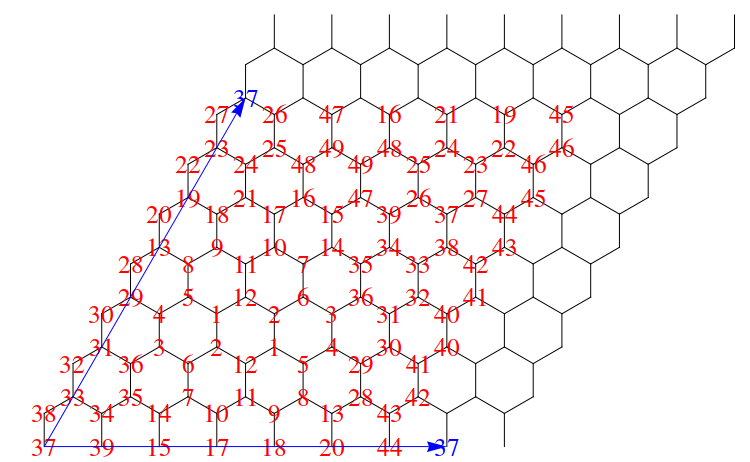}} \quad
\subfloat[]{\label{98structurefactor} \raisebox{16 pt}{\includegraphics[scale=0.3]{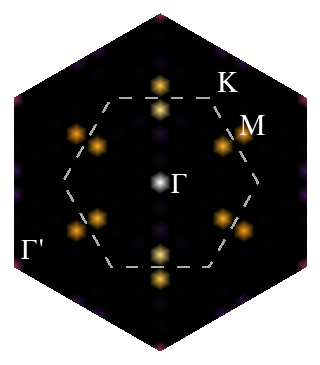}}}
\caption{\label{MOend}(a) The sublattice structure in the magnetic unit cell of the $98$-site order and the primitive vectors of the Bravais lattice it defines. For brievity, we do not show the real space spin configuration. (b) The static spin structure factor of the $98$-site order, which peaks at $\mathbf{k}=\Gamma, \frac{6}{7} \mathrm{M}_i$, with the parameters $K=-1$, $\Gamma=0.5$, $\Gamma'=-0.02$ and $h=0.6$. The intensities along the three fold directions are different, one direction is higher, while others are lower.}
\end{figure}

For each of the nontrivial magnetic orders, we show in Figs.~\ref{MOstart}-\ref{MOend} the sublattice structure in the magnetic unit cell, the real space spin configuration (except for the $70$-site and $98$-site orders) and the static spin structure factor
\begin{equation}
S_\mathbf{k} = \frac{1}{N_\mathrm{site}} \sum_{ij} \mathbf{S}_i \cdot \mathbf{S}_j e^{i \mathbf{k} \cdot (\mathbf{r}_i - \mathbf{r}_j)} ,
\end{equation}
where $\mathbf{r}_i=\mathbf{R}_i+sa/\sqrt{3} \hat{\mathbf{y}}$, $\mathbf{R}_i$ is the coordinates of the Bravais lattice, $a$ is the lattice constant, $s=0$ ($s=1$) if $\mathbf{r}_i$ belongs to the even (odd) sublattice of the honeycomb lattice and $N_\mathrm{site}$ is the total number of sites in the system. We choose the direct lattice vectors to be $\mathbf{a}_1=a \hat{\mathbf{x}}, \mathbf{a}_2= a (\cos \frac{\pi}{3} \hat{\mathbf{x}} + \sin \frac{\pi}{3} \hat{\mathbf{y}})$, so that the reciprocal lattice vectors are $\mathbf{b}_1=\frac{2 \pi}{a} \hat{\mathbf{x}} - \frac{2 \pi}{\sqrt{3} a} \hat{\mathbf{y}}, \mathbf{b}_2=\frac{4 \pi}{\sqrt{3} a} \hat{\mathbf{y}}$. The structure factors are plotted within the second Brillouin zone, and the first Brillouin zone is enclosed by dashed lines. Whenever our discussion involves more than one of the three inequivalent $\mathrm{M}$ points in the momentum space, we use the notation $\mathrm{M}_i, i=1,2,3$ to distinguish them. \\

We observe that the magnetic orders respect either an inversion $\mathcal{I}$ symmetry or a three fold rotational $C_3$ symmetry. Therefore, the magnetic orders can be divided into two classes. In Class $\mathcal{I}$, if two sites $i$ and $j$ are related by inversion, then the spins on these sites align exactly in the same direction. $i$ and $j$ belong to distinct sublattices of the honeycomb lattice, either $i$ even and $j$ odd, or $i$ odd and $j$ even. The inversion only acts in real space, i.e.~it does not flip the direction of the spin. In Class $C_3$, the spin configuration remains invariant under a $C_3$ rotation about the $[111]$ direction, with the axis of rotation piercing a site. The rotation takes place in both the real space and the spin space,
\begin{equation}
\mathbf{S}_i \overset{C_3}{\longrightarrow} C_3^{-1} \mathbf{S}_{C_3(i)} .
\end{equation}
In the spin space, $C_3$ permutes the $x$, $y$ and $z$ components of the spin. We illustrate the $\mathcal{I}$ and $C_3$ symmetries using the $18$-site and $18$-$C_3$ orders as examples in Figs.~\ref{18lowconfiguration} and \ref{18C3configuration}. In addition, each magnetic order in Class $\mathcal{I}$ has three different domains with the same energy, which are related by a $C_3$ rotation in both the real space and the spin space. Consequently, the profile of the structure factors corresponding to these three domains differs by a $C_3$ rotation. For clarity, however, there is no $C_3$ symmetry in each domain. Each magnetic order in Class $C_3$ has only one domain by definition. \\

Furthermore, we notice that there is perhaps some kind of number rule governing the size of the magnetic unit cells, which has yet to be understood. The $98$-site order appears like an augmented version of the $50$-site order, which in turn appears like an augmented version of the $18$-site order (see Figs.~\ref{18lowconfiguration} and \ref{50lowconfiguration}, for instance). Their respective magnetic unit cells contain $7 \times 7$, $ 5 \times 5$ and $3 \times 3$ unit cells of the honeycomb lattice. Besides, the $6$-site, $30$-site and $70$-site order, whose magnetic unit cells contain $3 \times 1$, $5 \times 3$ and $7 \times 5$ physical unit cells, appear like parts of the $18$-site, $50$-site and $98$-site orders, respectively. \\

\end{document}